\documentclass[a4paper,11pt]{article}
\usepackage{latexsym,amsfonts,amsmath,amssymb}
\usepackage{booktabs}
\usepackage[dvips]{graphicx}
\usepackage{subfigure, epsfig}
\usepackage{multicol}
\usepackage{fancyhdr}
\usepackage{float}
\usepackage[a4paper,left=3.0cm,right=3.0cm,top=3.5cm,bottom=3.0cm]{geometry}
\title{\bf Quantum entanglement of bound particles
under free center of mass dispersion}

\author{Fernanda Raquel Pinheiro$^{1,2,3}$\footnote{Electronic address: fep@fysik.su.se} and A. F. R. de Toledo Piza$^{1}$\\ \\
$^1$\it Instituto de F\'{i}sica, Universidade de S\~{a}o Paulo,  S\~{a}o Paulo, Brazil\\
$^2$\it Department of Physics, Stockholm University, 106 91 Stockholm, Sweden\\
$^3$\it NORDITA, 106 91, Stockholm, Sweden}
\date{}

\begin{document}
\maketitle

\begin{abstract}
  On the basis of the full analytical solution of the overall unitary dynamics,
the time evolution of entanglement is studied in a simple bipartite model system
evolving unitarily from a pure initial state. The system consists of two particles
  in one spacial dimension bound by harmonic forces and having its
  free center of mass initially localized in space in a minimum uncertainty wave packet. 
The existence of such initial states in which
  the bound particles are not entangled is pointed out. The
  entanglement of the two particles is shown to be independent of the
  wavepacket mean momentum, and to increase monotonically in a time
  scale distinct from that of the spreading of the center of mass
  wavepacket.

\end{abstract}

\section{Introduction}
%1 - Einstein Podolski Rosen
%2 - Schrodinger, 1935A
% descobrir referencias 3,4,5,6
%7 - Entanglement sudden death -> Davidovich

Since the early development of Quantum Mechanics, a number of questions
have been raised concerning its interpretation and status facing the
physical description of reality. `Entanglement' is one such question,
pointed out and developed by Schr\"{o}dinger~\cite{schrod} in the wake of
the paper by Einstein, Podolski and Rosen on the completeness of
Quantum Mechanics~\cite{einstein}. The situation considered by
Schr\"{o}dinger in that context hinges on the description of the quantum
mechanical state of each one of two separate parts of a composite
system, after they have interacted for some time. He observes that
even if, prior to the interaction, the state of each of the
constituent parts is described by its own quantum state vector (or, in
the words of Schr\"{o}dinger, by its {\it psi-function}), in which case
the quantum state of the composite system is described by
their product, this is in general no longer possible after interaction
has run its course, the constituent parts having become `entangled' by
the intervening interaction process. This entanglement is thus
associated to the non factorizability of the quantum state of the
composite system after the interaction, a trait of quantum mechanics
that ``enforces its entire departure from classical lines of
thought''~\cite{schrod}.

After the breakthrough of Bell~\cite{bell}, which opened the way to
experimental probing~\cite{aspect} into quantum peculiarities relating
to entanglement~\cite{einstein,mermin}, this concept has become a
subject of intense research~\cite{odradeks}, particularly in connection
with the fact that it provides crucial resources for the new research
domains of quantum information and quantum computation~\cite{niels,
  presk}. Prominent issues in this context have been, on the one side,
developing the characterization and quantification of entanglement,
especially as applicable to quantum states which are themselves not
represented in terms of a state vector, or a {\it
  psi-function}~\cite{odradeks,mint}; and, on the other side,
understanding the dynamics of entanglement specially in realistically
open systems, i.e. systems undergoing interactions with a complex
`environment'~\cite{mint}. Here one aims particularly at gaining
control over the degrading effects of external interactions on the
valuable entanglement properties residing in the system of interest.
These effects are generally referred to as `decoherence' processes, and
have been given also a positive role in connection with accounting for
the practical elusiveness of in principle allowed quantum
superpositions of macroscopically distinguishable
states~\cite{zurek,omnes}, i.e., in connection with the `construction
of a classical looking world' on the basis of quantum mechanics.

More recently, investigations of the effect of decoherence processes
on the time evolution of entanglement present in a quantum system led
to the finding of the rather surprising effect which became known as
ESD, ``entanglement sudden death'' (or, more euphemistically, ``early
stage disentanglement'')~\cite{yueb}, which consists in the complete
demise, within a finite time, of the entanglement initially present in
the system of interest. This effect has in fact been observed in terms
of a phenomenological effective, non unitary quantum dynamics which
was simulated with photon polarizations in the role of two level
systems~\cite{rio}. Subsequently ESD has been also found to be present
in extremely simple closed (tripartite) systems evolving
unitarily (in fact, multiperiodically) from initially entangled
psi-functions~\cite{lljn}, showing that environment induced decoherence
is not a requisite for this particular phenomenon in the dynamics of
entanglement. Furthermore, the fact that one of the subsystems
involved in the ESD process {\it does not interact} with the remaining
parts relativises the role of dynamical correlation processes in the
time evolution of entanglement.

This work deals with the dynamics of entanglement in the even simpler
context of a bipartite model system which is completely soluble in
analytical terms, using tools available in any introductory course on
quantum mechanics. The system consists of two particles in one spacial
dimension bound by harmonic forces and having its free center of mass
initially localized in space in a wave packet of minimum uncertainty.
By using center of mass and relative motion dynamical variables, the 
complete unitary quantum dynamical evolution of this state can be easily 
written down explicitly. Assuming that the two particles
are in a stationary state (e.g., the ground state, for simplicity) of
the relative motion, the non stationary character of the initial state
reduces therefore just to the dispersive time evolution of the center
of mass wavepacket. While the center of mass and relative `subsystems'
are initially not entangled, having their own psi-functions, and
remain so at all subsequent times, the same is in general not true
when one considers each of the two individual particles as subsystems.
It turns out, in fact, that the two particles are in general
entangled, and that their entanglement is time dependent as a result
of the dispersive time evolution of the center of mass wavepacket.
While the presence of entanglement is of no surprise, in view of the
fact that the two particles are strongly correlated dynamically by
being in a bound state of the harmonic interparticle force, one finds
nevertheless that the initial center of mass wavepacket can be
prepared so that the two particles are actually {\it not} entangled
initially. The dispersive time evolution of the center of mass
wavepacket is found to induce a monotonic increase in time of the
entanglement of the two particles, albeit in a time scale distinct from
that associated with the dispersive spreading of the center of mass
wavepacket. These facts exemplify in very simple terms the surprises
and subtleties one may find in the relation of entanglement of
subsystems and its dynamics to the unitary dynamical evolution of the
system as a whole.

The paper is organized as follows. In section 2 the complete
characterization of the model system is given and its dynamics is
fully worked out analytically, both in terms of center of mass and
relative coordinates and in terms of particle coordinates. The
presence of entanglement in the latter description is also discussed.
Section 3 constitutes the main body of the work. It is divided in
three sub-sections, dealing respectively with the description of the
quantum state of entangled subsystems, with some pertinent
quantitative measures of entanglement and with a rather detailed
discussion of the dynamics of entanglement in the model system.
Section 4 contains some concluding remarks. The explicitation of a
more technical point relating to the adopted measure of entanglement
is given in an Appendix, for completeness.

\section{Alternate descriptions of a flying two-body
  bound state}

We consider a system consisting of two (spinless) particles in just
one dimension, which interact through a translationally invariant
attractive potential. If the masses and positions of the particles are
$\{m_1, x_1\}$ and $\{m_2,x_2\}$ respectively and $V(x_2-x_1)$ is the
potential describing the interaction of the two particles, the
dynamics of the system is described by the Hamiltonian
\[
H = \frac{p_{1}^{2}}{2m_1} + \frac{p_2^2}{2m_2} + V(x_2-x_1),
\]
where $p_1$ and $p_2$ are the momenta canonically conjugated to $x_1$
and $x_2$. Note that the interaction potential $V$ correlates the
dynamics of the two particles. A common simplifying practice in this
context is to deal with this correlated two-body problem in terms of
an alternate set of dynamical variables describing the center of mass
and relative motions rather than in terms of the particle dynamical
variables themselves. The reason for this is that in terms of the new
variables one has only to deal with two {\it independent},
albeit effective, one-body problems. In fact, the appropriate
transformations to center of mass variables $X, P$ and relative
variables $x,p$ are
\begin{equation}
R = \frac{m_{1}x_{1} + m_{2}x_{2}}{m_1 + m_2},\hspace{.4cm}P=p_1+p_2;
\hspace{1cm}
r = x_2 - x_1,\hspace{.4cm} p = \frac{m_1p_2 - m_2p_1}{m_1 + m_2},
\label{transf}
\end{equation}
in terms of which the the Hamiltonian of the two-body problem appears
in the guise
\[
H = \frac{P^{2}}{2M}
+\left(\frac{p^{2}}{2\mu}+V(x)\right)\equiv H_{\rm CM}+H_{\rm rel}.
\]
Here the new mass parameters are, respectively, the total mass $M$ and the
reduced mass $\mu$ of the particles, which in terms of the particle masses
can be expressed as
\[
M = m_1 + m_2, \hspace{.4cm} \mu = \frac{m_1m_2}{m_1 + m_2}.
\]
This reduction procedure can be used both in classical and in quantum
contexts. In the former case it involves a canonical transformation,
while the latter it involves a unitary transformation of the basic
canonical variables. Even though this is in fact rather trivially
evident, it is worth noting explicitly that in both cases the system
maintains a {\it composite} character, but which in the latter
rendering reduces to two non-interacting parts.

We now recall the simple quantum mechanical treatment of the system,
taking as a specific instance of the two-body interaction potential a
simple harmonic potential $V(x)=\mu\omega^2x^2/2$. The stationary
solutions of the Schr\"{o}dinger equation for the composite system are
well known and simple to obtain. They consist in the product of a center
of mass momentum eigenfunction with a stationary state of the harmonic
oscillator equation which governs the dynamics of the relative motion.
Taking the relevant state to be the oscillator ground state one has
\[
\Psi(R,r) = e^{iKR}\frac{1}{\pi^{\frac{1}{4}}\sqrt{b}}e^{-\frac{1}{2} 
{(\frac{r}{b})}^{2}}, \hspace{1cm}b=\sqrt{\frac{\hbar}{\mu\omega}}. 
\]
Note that one faces here the well known normalization problem in the
case of the continuous spectrum of the center of mass Hamiltonian. We
will not have to deal explicitly with this, however, since we consider a
(non-stationary) normalized gaussian wavepacket of width $B$ and mean
momentum $\hbar K$ as the wavefunction describing the initial state of
the center of mass part. The relative part remains in its chosen
ground stationary state, so that our normalized non-stationary initial state 
can be written as
\begin{equation}
\Psi(R,r) =
\frac{1}{\pi^{\frac{1}{4}}\sqrt{B}}e^{-\frac{R^2}{2B^2}}
e^{iK.R}\frac{1}{\pi^{\frac{1}{4}}\sqrt{b}}e^{-\frac{1}{2}
  {(\frac{r}{b})}^{2}}.
\label{innorm}
\end{equation}

In order to obtain the time evolution of such a globally
non-stationary state it is most convenient to have it rewritten in the
momentum representation by means of a double Fourier transform
\begin{equation}
\Psi(\kappa,\xi) = \int e^{-i\kappa R} e^{-i\xi r}
\Psi(R, r)dR dr = 4\sqrt{\frac{Bb}{\pi}}\;\,e^{-2B^2(\kappa -
  K)^2}e^{-2{b}^2\xi^2}.
\label{instate}
\end{equation}
Since the relative state is an oscillator eigenstate, the Fourier
transform involving the relative coordinate is strictly not necessary
for this purpose, but it will be convenient further on for reverting
the description to the original particle variables, instead of the
center of mass and relative variables.

Time evolution can now be implemented by just bringing in the
appropriate time-dependent phase factors associated with the different
center of mass momentum eigenfunctions as well as with the relative
oscillator eigenfunction. This leads to the well known translation and
spreading of the center of mass wave packet, with no other effect than
an irrelevant overall time-dependent phase factor originating from the
internal energy eigenstate. More formally, the time evolution is
represented by the action of the evolution operator 
\[
U(t,0)=e^{-\frac{i}{\hbar}\frac{P^2t}{2M}}\otimes 
e^{-\frac{i}{\hbar}H_{\rm    rel}t},
\]
which reduces to numerical phase factors when acting on
the momentum space representation. Thus, after elapsed time $t$, the
momentum space wave-function (\ref{instate}) becomes
\begin{equation}
\Psi(\kappa, \xi, t) = 4\sqrt{\frac{Bb}{\pi}}\;\, e^{-\frac{i\hbar
    \kappa^2t}{2M}}e^{-2B^2(\kappa - K)^2}\;\,e^{-i\frac{\omega}{2}t}
e^{-2{b}^2\xi^2}.
\label{tevstate}
\end{equation}

The simplicity of this result is basically due to the independence of
the center of mass and relative dynamics. In particular, the
stationary character of the relative motion state implies that its
contribution to the joint time-evolution reduces to an overall time
dependent phase factor. The time evolution of the center of mass wave
packet is itself more involved, as different momentum components
acquire different time-dependent phases which govern its dispersive
displacement. The center of mass and relative motion parts of
the composite system have however their independence fully
uncompromised by these time-dependences, as clearly exhibited by the
factorization of the overall wavefunction into center of mass and
relative motion factors {\it at all times}.

Now let us return to our former decomposition of the system, in terms of
its two constituent particles. In momentum representation, this can be done on the 
basis of the solution written in eq. (\ref{tevstate}), through reverting to the dynamical
variables relating to the individual particles. The appropriate transformation here is 
(cf. eqs. (\ref{transf}))
\[
\kappa = k_1 + k_2,\hspace{.4cm}\xi = \frac{m_1k_2 - m_2k_1}{M},
\]
where $k_1$ and $k_2$ are the wave numbers associated with the
momentum eigenfunctions of particles 1 and 2 respectively. In terms of
these, the time evolved state appears as
\begin{equation}
\Psi(k_1,k_2,t)=4\sqrt{\frac{Bb}{\pi}}\;\, e^{\frac{-i\hbar t}{2M}
(k_1+k_2)^2}e^{-2B^2(k_1+k_2 -
K)^2}e^{-i\frac{\omega}{2}t}e^{-\frac{2b^2}
{M^2}(m_1k_2-m_2k_1)^2}.
\label{tev2part}
\end{equation}

An important feature of this alternate rendering of the state of the
two-particle system is that the dependence on $k_1$ and on $k_2$ does
not factor {\it for all times $t>0$}, as can be seen by the occurrence
of terms involving the product $k_1k_2$ in the exponents. Note however
that factorization will in fact occur at $t=0$ for a particular choice
of width parameter $B$ of the center of mass wavepacket. In fact, when
this parameter is tuned to the value satisfying

\begin{equation}
B^2\rightarrow B_0^2=\frac{m_1m_2}{M^2}b^2=\frac{\mu}{M}b^2
\label{fact}
\end{equation}

\noindent these non-factoring terms drop out, and one obtains a
product wavefunction also in terms of the individual particle
variables. This is clearly a result particular to our choice of a real
gaussian envelope function for the center of mass wavepacket, combined
with the gaussian form of the relative ground state in the harmonic
particle-particle interaction potential.

The factorizability of the wavefunction (\ref{tevstate}) {\it at all
  times $t$} allows for associating the quantum state of the center of
mass and relative parts of the system with a respective factor of that
wavefunction. Each of these two parts is therefore endowed with its
own independent quantum mechanical probability amplitude. Non
factorizability when it is expressed in terms of particle variables as
in eq. (\ref{tev2part}), on the other hand, implies that the joint
probability amplitude includes such correlation involving the two
particles which impedes an independent description of each one of them
in terms of probability amplitudes. This is just what has been called
by Schr\"{o}dinger (in 1935, \cite{schrod}) the quantum `entanglement' of
the two particles.

Our findings so far indicate moreover that a definite distinction
should be made between entanglement and correlations between parts of
the composite system. In fact, in the state described by eqs.
(\ref{tevstate}) and (\ref{tev2part}), the two particles are, at all
times, and in particular at $t=0$, strongly correlated by finding
themselves bound in the stationary ground state of relative motion.
The initial state factorization in terms of particle variables for
$B=B_0$ provides thus for a counter example disallowing direct
association of such correlation with entanglement. The factorized
amplitudes of the wavefunction cast in terms of the particle
variables, eq.  (\ref{tev2part}), in a generic factorized instance
$B^2=\mu b^2/M$ at $t=0$, appear as

\[
\psi_1(k_1)=N_1e^{-2b^2\frac{\mu}{M}(1+\frac{m_2}{m_1})k_1^2+4b^2\frac{\mu}
{M}k_1K}\hspace{.5cm}{\rm and}\hspace{.5cm}\psi_2(k_2)=N_2e^{-2b^2
\frac{\mu}{M}(1+\frac{m_1}{m_2})k_2^2+4b^2\frac{\mu}{M}k_2K},
\]

\noindent $N_1$ and $N_2$ being the appropriate normalization
constants. It is apparent in these expressions that, while the {\it
  quantum kinematical} property of amplitude factorization ensures
lack of entanglement, each one of the two factor amplitudes involve
parameters referring to the two particles and bearing on their
dynamical correlation.

\section{Quantum states of entangled particles}

According to the rules of quantum mechanics, knowledge of the joint
wavefunction (\ref{tev2part}) provides us with the most complete
information one can obtain about the system itself. It must then, in
particular, also provide us with the most complete information one can
obtain about each of its constituent particles, considered separately.
The formal tool usually employed to extract these more restricted
pieces of information from the amplitude (\ref{tev2part}) describing
the quantum state of the two-particle system involves first setting up
the so called {\it density matrix}, defined as

\begin{equation}
\rho(k_1,k_2;k_1',k_2';t)\equiv\Psi(k_1,k_2,t)\Psi^*(k_1',k_2',t).
\label{dm}
\end{equation}

\noindent The `diagonal' part of this, i.e., the part with $k_1=k_1'$
and $k_2=k_2'$, corresponds just to the joint probability distribution
(in momentum space) of the two particle system, at time
$t$. Moreover, from the normalization of the wavefunction
$\Psi(k_1,k_2,t)$ it follows that

\[
\int dk_1''\int dk_2'' \rho(k_1,k_2;k_1'',k_2'';t)
\rho(k_1'',k_2'';k_1',k_2';t)= \rho(k_1,k_2;k_1',k_2';t).
\]

\noindent The double integration works here just as a matrix
multiplication, so that this relation in fact shows that, considered
as representing an operator, the density matrix (\ref{dm}) associated
with the normalized wavefunction (\ref{tev2part}) is hermitian
and {\it idempotent}, and therefore a projection operator.  

Now the mean value of {\it any} observable ${\cal{O}}^{(1)}$
pertaining to the first particle can be obtained in terms of the
density matrix as

\begin{eqnarray*}
\langle{\cal{O}}^{(1)}\rangle_t&=&\int dk_1\int dk_1'
{\cal{O}}^{(1)}_{k_1'k_1}\int dk_2\;\rho(k_1,k_2;k_1',k_2;t)\equiv \\
&\equiv&\int dk_1\int dk_1'{\cal{O}}^{(1)}_{k_1'k_1}\bar{\rho}^{(1)}
(k_1,k_1';t),
\end{eqnarray*}

\noindent where ${\cal{O}}^{(1)}_{k_1'k_1}$ is the momentum
representation of the observable under consideration,
${\cal{O}}^{(1)}$. Since this works for any observable pertaining to
particle 1, it follows that the object

\[
\bar{\rho}^{(1)}(k_1,k_1';t)\equiv\int dk_2\;\rho(k_1,k_2;k_1',k_2;t),
\]

\noindent known as the {\it reduced density matrix} for particle 1
contains all information concerning this particle, when considered
independently of the rest of the system. The operation leading to the
reduced density matrix is usually referred to as {\it taking a partial
  trace} of the density matrix $\rho(k_1,k_2;k_1' ,k_2';t)$ over
particle 2, and amounts to an all inclusive acceptance of whatever
properties it would have correlated to the particle under scrutiny. A
similar argument allows one to collect all information concerning
particle 2 which is similarly inclusive with respect to properties
of particle 1, by taking a partial trace of $\rho(k_1,k_2;k_1'
,k_2';t)$ over this particle, i.e.

\[
\bar{\rho}^{(2)}(k_2,k_2';t)\equiv\int dk_1\;\rho(k_1,k_2;k_1,k_2';t).
\]

\noindent Note that, again in view of the normalization of the
wavefunction (\ref{tev2part}), one has

\begin{equation}
\int dk_1\bar{\rho}^{(1)}(k_1,k_1;t)=\int dk_2\bar{\rho}^{(2)}
(k_2,k_2;t)=1.
\label{trace1}
\end{equation}

An application of these ideas to the specific case of the
two-particle system as considered here goes as follows. First, the
density matrix reads explicitly (cf. eqs. (\ref{tev2part}) and
(\ref{dm}))

\begin{eqnarray}
\rho(k_1,k_1',k_2,k_2',t)&=& \frac{16Bb}{\pi}\;\,e^{\frac{-i\hbar t}{2M}
\left((k_1+ k_2)^2 -(k_1'+ k_2')^2)\right)}
e^{-2B^2\left((k_1+k_2 - K)^2+(k_1'+ k_2'-K)^2\right)}\times\nonumber\\
&&\times e^{-\frac{2b^2}{M^2}\left((m_1k_2 - m_2k_1)^2 + 
(m_1k_2' - m_2k_1')^2\right)}.     
\label{fulldens}
\end{eqnarray}

\noindent Next, the reduced density matrix for particle 1 can be
obtained by taking a partial trace of this over particle 2, which
amounts here to setting $k_2=k_2'$ and performing a gaussian integration
over the momentum variable $k_2$. This gives the somewhat lengthy result

\begin{eqnarray}
\label{redens1}
&&\hspace{-1.2cm}\rho^{(1)}(k_1,k_1';t)= \\
&=& \frac{8BbM\;\, e^{-\frac{4K^2B^2b^2m_1^2}{M^2B^2 +
    m_1^2b^2}}}{\pi\sqrt{M^2B^2+m_1^2b^2}}\;\, 
e^{-\frac{\hbar^2t^2}{16(M^2B^2 + m_1^2b^2)}(k_1 -
  k_1')^2}e^{4KB^2(k_1 - k_1')}e^{\frac{(\mu b^2 - MB^2)^2}{M^2B^2 +
    m_1^2b^2}(k_1 + k_1')^2}\times \nonumber \\
&\times& e^{2KB^2M\left(\frac{\mu b^2 - MB^2}{M^2B^2
    + m_1^2b^2}\right)(k_1 + k_1')}
e^{-\frac{2}{M^2}(M^2B^2 + m_2^2b^2)(k_1^2 + k_1'^2)}
e^{-\frac{i\hbar t}{2}\left( \frac{m_1b^2}{M^2B^2 + m_1^2b^2}\right)
(k_1^2 -  k_1'^2)}. \nonumber
\end{eqnarray}

\noindent In view of the complete symmetry of the density matrix
(\ref{fulldens}) under an interchange of indices $1\leftrightarrow 2$,
the reduced density matrix for particle 2 can be immediately read off
from eq.  (\ref{redens1}) simply by replacing all indices 1 by 2 and
vice versa.

\subsection{Reduced density factorization and entanglement} 

An important feature of the reduced density matrix (\ref{redens1}) and
also of its companion $\rho^{(2)}(k_2,k_2';t)$ is that their
dependence on primed and unprimed variables does not factor in
general. This is in fact clear in view of the exponentials involving
$(k_i\pm k_i')^2$. Non-factorizability implies that the reduced
densities cannot be obtained from single quantum amplitudes in the
same way that the full density matrix (\ref{dm}) is obtained from the
wavefunction (\ref{tev2part}), and indicates the entanglement of the
two particle system in the selected quantum state\footnote{Recall that
  factorization may occur in special cases, such as the one pointed
  out earlier (see eq. (\ref{fact})) in which the wavefunction
  (\ref{tev2part}) factors, namely $t=0$, and $B\rightarrow
  B_0=b\sqrt{\mu/M}$. In this particular situation it is possible to
  associate probability amplitudes to each of the two particles,
  although each of the amplitudes will involve the mass parameters of
  both particles, in addition to `shared' parameters such as the
  oscillator parameter $b$ and the width $B_0$ of the center of mass
  wavepacket.}. The reduced densities are however hermitian objects in
the sense that

\[
\rho^{(i)*}(k_i,k_i';t)=\rho^{(i)}(k_i',k_i;t),\hspace{1cm}i=1,2\;\,.
\]

\noindent A useful representation of the reduced densities can
therefore be obtained in terms of the solution of the eigenvalue
problem

\[
\int\rho^{(i)}(k_i,k_i';t)\chi^{(i)}_n(k_i',t)\,dk_i'=\lambda^{(i)}_n
(t)\chi^{(i)}_n(k_i,t).
\]

\noindent Hermiticity ensures that the eigenvalues are real and the
eigenfunctions can be made orthonormal. In addition, from the fact
that the mother density (\ref{dm}) is a projection operator, one can
actually deduce that the eigenvalues are in fact {\it non negative}, i.e.
$\lambda^{(i)}_n(t)\geq 0$.  The reduced density $\rho^{(i)}(k_i,k_i';t)$
can therefore be written in the form

\begin{equation}
\rho^{(i)}(k_i,k_i';t)=\sum_n\lambda^{(i)}_n(t)\;\chi^{(i)}_n(k_i,t)
\chi^{(i)*}_n(k_i',t)
\label{specrepi}
\end{equation}

\noindent which engages all the (orthonormal) eigenfunctions with
nonvanishing eigenvalue. Note also that, in view of the normalization
property (\ref{trace1}), one has $\sum_n\lambda^{(i)}_n(t)=1$.

A rather non trivial outcome of this representation is that the two
reduced densities originating from a given mother density matrix of
the form (\ref{dm}) {\it have the same eigenvalues}, their
eigenfunctions being related as

\[
\chi^{(2)}_n(k_2,t)=N_n(t)\int dk_1\chi^{(1)*}_n(k_1,t)\Psi(k_1,k_2,t)
\]

\noindent where $N_n(t)$ is an appropriate normalization factor.
Furthermore, the joint wavefunction itself may be written in terms of
the reduced densities eigenfunctions as (see Appendix))

\begin{equation}
\Psi(k_1,k_2,t)=\sum_n\sqrt{\lambda_n(t)}\chi^{(1)}_n(k_1,t)
\chi^{(2)}_n(k_2,t),
\label{schmidt}
\end{equation}

\noindent in what is usually referred to as the Schmidt decomposition~\cite{presk}.

Therefore, what one can do in general is to write the reduced
densities as linear combinations of orthogonal projection operators
with non negative coefficients of unit sum, also known as {\it convex
  sums} of orthogonal projection operators. Note that cases in which
dependence on primed and unprimed variables factors in the reduced
densities are duly covered as cases in which a single nonvanishing,
hence unit valued, eigenvalue occurs in eq.  (\ref{specrepi}). This
makes the reduced densities themselves projection operators.

Whenever the convex sum (\ref{specrepi}) involves more than a single
term the same happens with eq. (\ref{schmidt}), which reveals a
connection of non factorization of the dependence on primed and
unprimed variables in the reduced densities and the non factorization
of the dependence on $k_1$ and $k_2$ in the wavefunction of the
composite system. This is precisely the technical characterization of
entanglement given by Schr\"{o}dinger in his paper of 1935~\cite{schrod}.
Thus, entanglement is revealed in this case by the fact that the
reduced densities have more than a single nonvanishing eigenvalue
(hence needing more than a single eigenfunction for its
representation). In view of eq. (\ref{schmidt}) one can see this as
scars left from the process of `orphaning' of quantum amplitude
correlations implied by multiple terms in eq. (\ref{schmidt}).

Finally, it is worth stressing that the preceding analysis has been
performed on objects ultimately related to the wavefunction of the
composite system {\it at a given time $t$}. Therefore all the derived
quantities are in general time dependent in a way which can be
ultimately traced back to the unitary time evolution of the
wavefunction of the composite system. In particular, if one devises a
quantitative measure of entanglement, one will be able to follow its
evolution in time.

\subsection{A simple quantitative measure of entanglement}

The coefficients in the convex sum over orthogonal projection
operators in eq. (\ref{specrepi}), given the condition
$\sum_n\lambda_n(t)=1$ which follows from normalization of the
wavefunction of the composite system, can be seen as a probability
distribution over the set of projections onto the eigenvectors of the
reduced density. Furthermore, as the occurrence of more than a single
nonvanishing weight signals entanglement of the two parts of the
considered system which are under consideration, one is led to the use
of entropy-like functions as they are used in statistical mechanics to
provide for quantification of the degree of entanglement. A common
choice for that is the so called von Neumann
entropy~\cite{vonNeumann,odradeks}, given as

\[
S[\rho^{(1)}(t)]=-\sum_n\lambda_n\;\log\lambda_n.
\]

\noindent This is a function endowed both with additivity and convexity
properties\cite{Falk}, but its evaluation is often more costly than
alternate functions, such as the so called `2-R\'{e}nyi' entropy $S_{2R}$
and the `linear' entropy $\Delta$~\cite{odradeks}, which are defined
respectively as

\[
S_{2R}[\rho^{(1)}(t)]=-\log\sum_n\lambda_n^2\hspace{.5cm}{\rm and}
\hspace{.5cm}\Delta[\rho^{(1)}(t)]=\sum_n\lambda_n(1-\lambda_n)=
1-\sum_n\lambda_n^2.  
\]

\noindent The evaluation of the sum appearing in these objects can be
carried out in a simple way in terms the reduced density matrix itself, by
noting that

\[
\sum_n\lambda_n^2={\rm Tr}[\rho^{(1)}(t)]^2=\int dk_1\int dk_2
\;\rho^{(1)}(k_1,k_2;t)\rho^{(1)}(k_2,k_1;t).
\]

\noindent An identical result is obtained if the reduced density
$\rho^{(2)}(k_1,k_2;t)$ is used instead of $\rho^{(1)}(k_1,k_2;t)$. 

Both of these functions are non-negative and have the desirable
convexity properties, although the linear entropy is {\it not
  additive}. This fact is however of little relevance in this context,
and since, unlike the values of $S_{2R}$, the values of $\Delta$ are
bounded above by 1, we will use the linear entropy for quantifying the
entanglement of the two particles.

Evaluation of the linear entropy for the reduced density obtained in
eq. (\ref{redens1}) gives

\begin{eqnarray}
\Delta(t)=1 -{\rm Tr}[\rho^{(1)}(t)]^2&=& 
1 - \frac{1}{\sqrt{\frac{\hbar^2t^2}{16M^2B^2b^2} +
    \frac{(M^2B^2 + m_1^2b^2)(M^2B^2 + m_2^2b^2)}{M^4B^2b^2}}}
\equiv \nonumber \\ &\equiv& 1-\frac{b}{B}\frac{\tau_B}{\tau}
\frac{1}{\sqrt{1+\frac{t^2}{\tau^2}}}\;\,.
\label{delta}
\end{eqnarray}

\noindent In the last expression, $\tau_B=MB^2/\hbar$ is the
characteristic time associated with the dispersive spreading of the
center of mass wavepacket, and $\tau$ has been introduced as the
characteristic time associated with the time dependence of the linear
entropy. It is given by

\begin{equation}
\tau\equiv\tau_B\sqrt{\left(1+\frac{m_1^2b^2}
{M^2B^2}\right)\left(1+\frac{m_2^2b^2}{M^2B^2}\right)}\;,
\label{tau}
\end{equation}

\noindent an expression which is symmetric under the interchange of
indices $1\leftrightarrow 2$, consistently with the fact that the same
result for the linear entropy is obtained if one uses
$\rho^{(2)}(k_2,k_2',t)$, instead of $\rho^{(1)}(k_1,k_1',t)$, in its
evaluation.

\subsection{Initial entanglement and its dynamics}

A first, general feature to be noted in connection with the expression
obtained for the linear entropy $\Delta(t)$ is that it turns out to be
independent of the mean center of mass momentum $K$. This indicates
Galilean invariance of the dynamics of entanglement in the present
case, and is similar to what happens in connection with the dispersive
spreading of wavepackets in position space. Actually this feature can
be examined in very simple terms by considering the action of the
center of mass momentum translation operator

\[
G(\zeta)= e^{i\zeta X} = e^{i\zeta\frac{m_1x_1+m_2x_2}{M}}
\]

\noindent on the representation of the wavefunction of the composite
system given in eq. (\ref{schmidt}). The fact that the two position
operators $x_1$ and $x_2$ commute allows for the factorization of
$G(\zeta)$ as $G_1(\zeta)\otimes G_2(\zeta)$, so that

\[
G(\zeta)\Psi(k_1,k_2,t)= \sum_n \sqrt{\lambda_n} \left(G_1(\zeta)
\chi_n^{(1)}(k_1,t)\right)\left(G_2(\zeta)\chi_n^{(2)}(k_2,t)\right).
\]

\noindent The center of mass momentum translation therefore does not
affect the amplitudes $\sqrt{\lambda_n}$ (also known as the coefficients of 
the Schmidt decomposition) which govern the
entanglement. The eingenvalues of the reduced densities are therefore
preserved, and so is, in particular, the value of the linear entropy.

\begin{figure}
\centering
\includegraphics[width=4.5in]{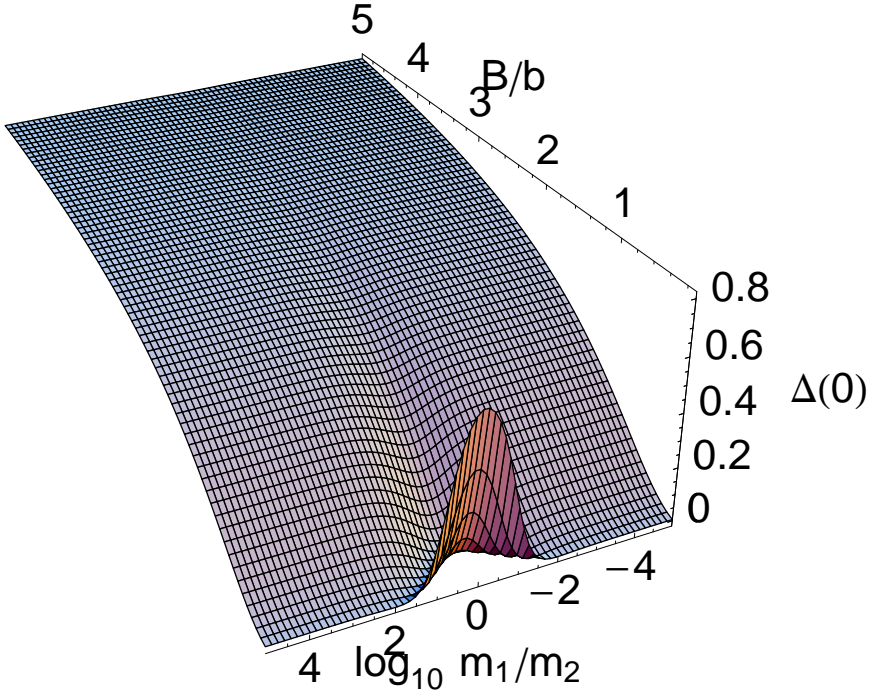}
\caption{\small Overall view of the initial linear entropy landscape
  as a function of $\log_{10} m_1/m_2$ and
  $B/b$. The logarithmic scale for the mass ratio
  emphasizes the symmetry under interchange of $m_1$ and $m_2$. Most
  interesting features are seen to develop along the plane $m_1/m_2=1$
  ($\log_{10}m_1/m_2=0$), mainly around $B/b=0.5$.}
\label{bigview}
\end{figure}

Consider next the value of the reduced entropy for the class of
initial states of the form (\ref{instate}). It is given in terms of
the the mass parameters $m_1$, $m_2$ and of the gaussian width
parameters $b$, $B$ as

\begin{eqnarray}
\Delta_0\equiv\Delta(t=0)=1-\frac{b}{B}\frac{\tau_B}{\tau}&=&
1-\frac{b}{B}\left[\left(1+\frac{m_1^2b^2}{M^2B^2}\right)\left(1+
\frac{m_2^2b^2}{M^2B^2}\right)\right]^{-\frac{1}{2}}= \nonumber \\
&=&1-\left[\frac{B^2}{b^2}\left(1-\frac{b^2}{B^2}\frac{m_1/m_2}
{(1+m_i/m_2)^2}\right)^2+1\right]^{-\frac{1}{2}}.
\label{linent0}
\end{eqnarray}

\noindent As shown, it can be expressed in terms of just the two
dimensionless quantities $B/b$ and $m_1/m_2$. Their value,
supplemented by one additional scale setting quantity, which can be
conveniently taken to be the total mass $M$, completely determines
particular initial states of the assumed form. Note that the linear
entropy is independent of the scale setting, so that its behavior
over different domains in the manifold of states of the form
(\ref{instate}) may be mapped by analyzing $\Delta_0$ as a function of
$B/b$ and $m_1/m_2$. An overall view of the behavior of this function
can be seen as a surface plot in fig. \ref{bigview}.

We now discuss some features of this initial entanglement landscape
which can be scrutinized by analyzing eq. (\ref{linent0}) in some detail.
First of all, one easily finds that, for any given value of the mass
ratio $m_1/m_2$, the relevant extremum of $\Delta_0$ as a function of
$B/b$ occurs for $B/b\rightarrow\sqrt{m_1m_2}/(1+m_1/m_2)$.
Substitution back into the expression for $\Delta_0$ shows that one
has at these points $\Delta_0=0$. We thus re-obtain, in the guise
of the vanishing of the initial value of the linear entropy, the
parametric conditions for the factorization of the initial
wavefunction when cast in terms of particle variables.

\begin{figure}
\centering
\includegraphics[width=4in]{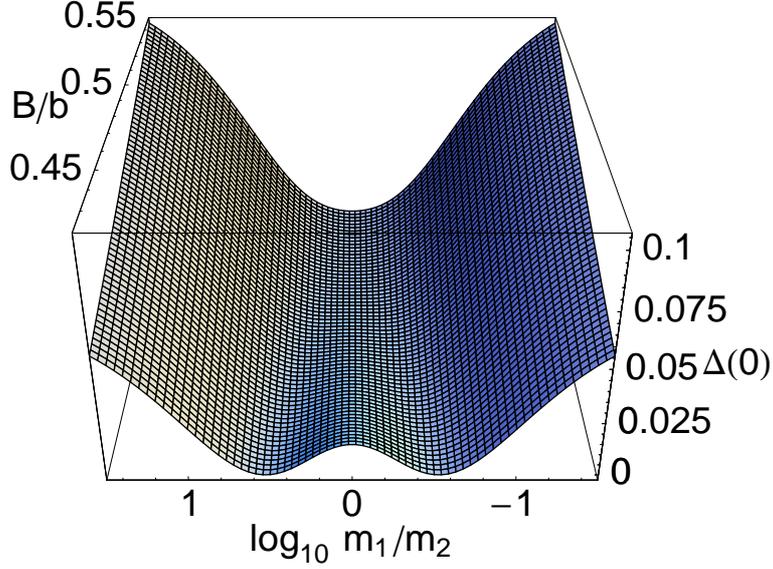}
\caption{\small Closer view of the most interesting section of the
  initial linear entropy landscape as a function of $\log_{10}
  m_1/m_2$ and $B/b$. The section shown here contains the point
  $m_1/m_2=1$, $B/b=0.5$ where the valley of $\Delta(t=0)$ at
  $m_1/m_2=1$ for $B/b>0.5$ branches into two valleys where
  $\Delta(0)=0$, separated by a ridge still at $m_1/m_2=1$, for
  $B/b<0.5$.}
\label{Delta3D}
\end{figure}

\begin{figure}
\centering
\includegraphics[width=4in]{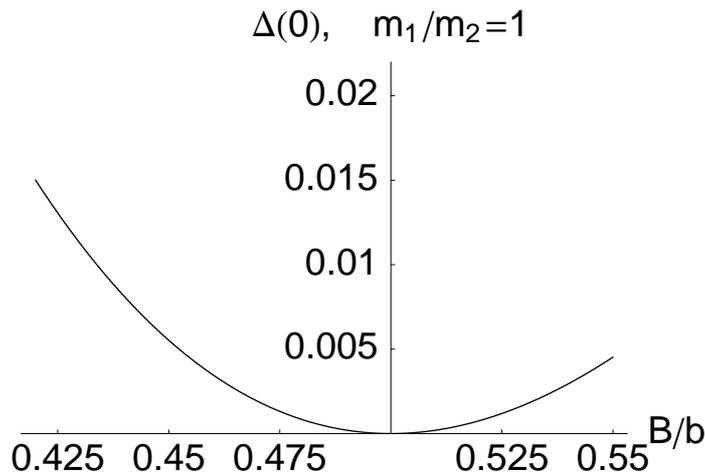}
\caption{\small A cut of the surface shown in fig. \ref{Delta3D} at
  the plane $\alpha=m_1/m_2=1$, showing values of the maxima (for
  $\beta<0.5$) and minima (for $\beta>0.5$) of the initial linear
  entropy. The plotted range of $\beta=B/b$ is the same as in fig.
  \ref{Delta3D}.}
\label{Delta1}
\end{figure}

All the possible values of $B/b$ associated with this set of
unentangled initial states lie in the interval $0\leq B/b\leq 0.5$, as
follows from the range available for $m_1/m_2$, namely $0\leq
m_1/m_2<\infty$. They are therefore confined to a front vertical slice
of fig. \ref{bigview}. In order to explore the behavior of the initial
value of the linear entropy for $B/b>0.5$ one may fix the value of
this variable and study the dependence of the initial linear entropy
on $m_1/m_2$. What one finds in this way is that there is an extremum
at $m_1/m_2=1$ for all possible values of $B/b$, in addition to two
other extrema with $m_1/m_2$ values given by

\[
\left(\frac{m_1}{m_2}\right)_\pm=\frac{b^2}{2B^2}
\left(1\pm\sqrt{1-4B^2/b^2}\right)-1
\]

\noindent which are real, non-negative numbers only for $B/b\leq 0.5$.
It turns out that the extremum at $m_1/m_2=1$ is in fact a
(nonvanishing) minimum of the initial linear entropy for $B/b>0.5$,
which however becomes a maximum for $B/b<0.5$. In this range, the
additional extrema are the minima for which $\Delta(0)=0$, which have
already been identified earlier. The fact that there are {\it two} such minima
is due to the symmetry of the linear entropy under interchange of the
masses $m_1$ and $m_2$. As a check on this, one can easily verify that
$(m_1/m_2)_-(m_1/m_2)_+=1$ for all values of $B/b$. At $B/b=0.5$,
$\Delta(0)$ has a very flat minimum (a zero of the fourth order) as a
function of $m_1/m_2$ at $m_1/m_2=1$. These features are displayed in
fig. \ref{Delta3D}, which covers the relevant domain of the initial
linear entropy landscape, namely $0.031\leq m_1/m_2\leq 31$ and
$0.42\leq B/b\leq 0.55$. A cut of this surface graph, showing the
values of $\Delta(0)$ at $m_1/m_2=1$ as a function of $B/b$ is shown
in fig. \ref{Delta1}. In this graph the ranges $B/b<0.5$ and $B/b<0.5$
correspond respectively to a ridge and to a valley in the $\Delta(0)$
surface.
  
\begin{figure}
\centering
\includegraphics[width=4.5in]{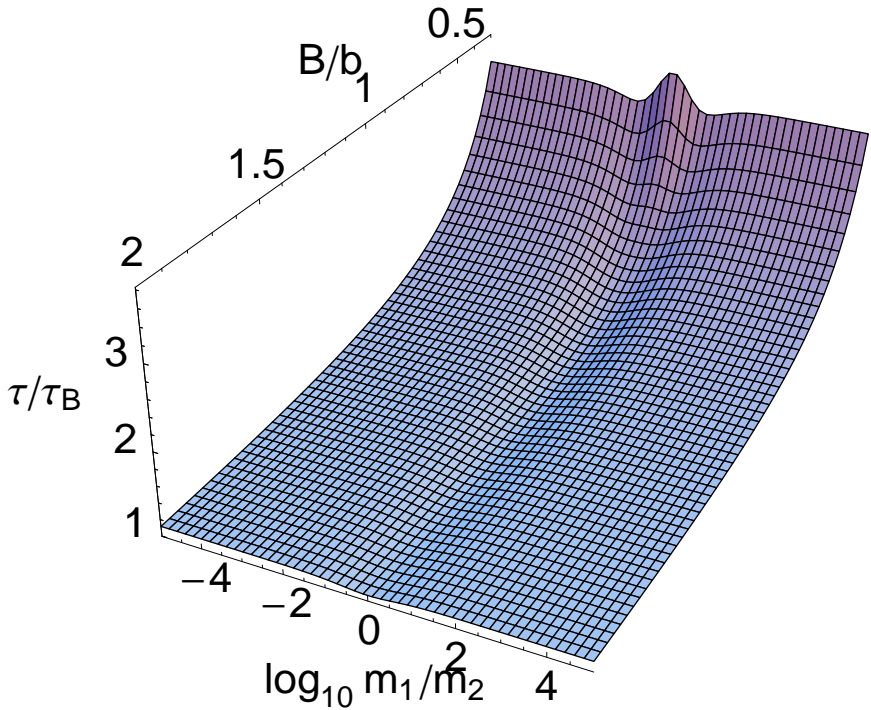}
\caption{\small Overall view of the $\tau/\tau_B$ landscape as a
  function of $\log_{10} m_1/m_2$ and $B/b$. Note the relation to the
  features of the corresponding landscape for the initial linear
  entropy.}
\label{bigtau}
\end{figure}

The time dependence of the linear entropy measure of particle-particle
entanglement consists in a simple, monotonic approach to the upper
bound $\Delta=1$ as seen in eq. (\ref{delta}). It involves an
algebraic expression which is of the same form as that which governs
the dispersive spreading of the center of mass wavepacket albeit
involving a time scale $\tau\neq \tau_B$. Since $\tau_B$ is the
relevant time scale in the unitary dynamical evolution of the initial
state of the composite system, it is both natural and convenient to
use this time scale to analyze the dependence of the characteristic
time $\tau$ of the dynamics of entanglement on the different initial
states of the form (\ref{instate}).

The quantity $\tau/\tau_B$ given in eq. (\ref{linent0}) is seen to be
closely related to the initial state value of the linear entropy. In
particular, it is also independent of the scale setting parameter $M$,
and can be written explicitly as

\begin{equation}
\frac{\tau}{\tau_B}=\frac{b/B}{1-\Delta(0)}=\left[\left(1-\frac{b^2}
{B^2}\frac{m_1/m_2}{(1+m_1/m_2)^2}\right)^2+\frac{b^2}{B^2}\right].
\label{ttb}
\end{equation}

\begin{figure}
\centering
\includegraphics[width=4in]{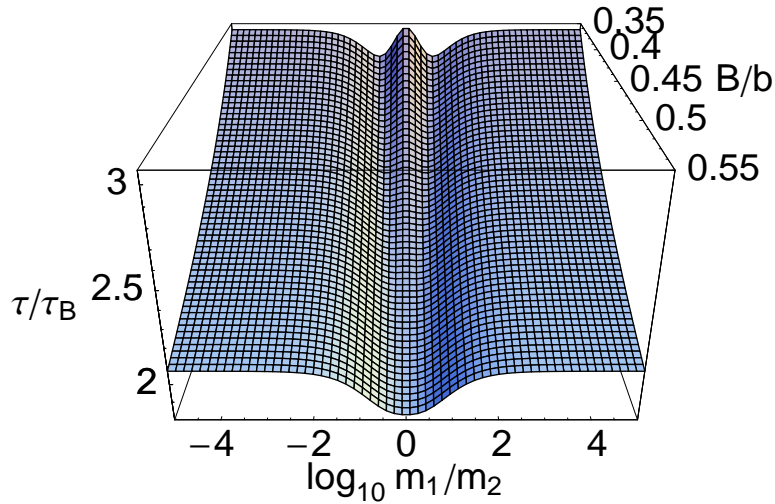}
\caption{\small Closer view of the valley bifurcation at $m_1/m_2=1$,
  $B/b=0.5$ in the $\tau/\tau_B$ landscape. Compare with
  fig. \ref{Delta3D}.} 
\label{tau3D}
\end{figure}

\noindent This implies that the main general features of the
$\Delta(0)$ landscape are essentially carried over to the
$\tau/\tau_B$ landscape, as can in fact be clearly in fig.
\ref{bigtau}. One finds here the same overall structure of valleys and
ridges seen in the corresponding initial linear entropy plot, in fig.
\ref{bigview}. A similar comparison can be made for the close-ups into
the more intricate region of the $m_1/m_2\times B/b$ plane, figs.
\ref{tau3D} and \ref{Delta3D}. As a consequence of inversion and of an
additional $b/B$ factor, however, one sees that $\tau/\tau_B$ is
bounded by unity {\it below}, while $\Delta(0)$ is similarly bounded
{\it above}. For sufficiently small values of $B/b$ the characteristic
time for the evolution of entanglement becomes substantially longer
than $\tau_B$.

Note however that more tightly localized wavepackets spread faster
(have smaller values of $\tau_B$), so that this does not imply very
long `absolute' times for entanglement evolution.  Using the value of
$M$ for scale setting, the reduced mass $\mu$ and the value of $b$ are
set by the values of $m_1/m_2$ and of the oscillator frequency
$\omega$, the wavepacket width $B$ being finally fixed by $B/b$. The
limiting value of the `absolute' characteristic time $\tau$ for
tightly confined center of mass wavepackets fixed in this way is

\[
\lim_{B\rightarrow 0}\tau=\frac{\mu b^2}{\hbar},
\]

\noindent which is just the inverse of the oscillator frequency
$\omega$.

\section{Concluding remarks}

We have examined in some detail the quantum entanglement of two
particles in one dimension whose center of mass evolves as a free
gaussian wavepacket with mean momentum $\hbar K$, when they are bound
in the ground state of a two-body harmonic interaction
potential. While the fact that the occurrence of entanglement is of
no surprise in view of the strong mutual correlation (e.g. in position
space) of the two particles~\cite{PRIT}, one can easily see in this
case that the entanglement may nevertheless be made to vanish in the
initial state by an appropriate choice of the width parameter of the
center of mass wavepacket. The fact that particle-particle
correlations can, at least in some cases, be represented in unentagled
form (which, in the present case, means being represented by
factorized amplitudes) indicates that the two concepts are in fact
quite distinct.

Given the stationary character of the bound state of the two
particles, the dynamics of their entanglement is driven by the
unitary, dispersive spreading of the center of mass wavepacket. In
particular, it also comes out as being independent of the mean value
of the center of mass momentum, a fact which can be understood in
terms of the preservation of the coefficients in the Schmidt
decomposition of the two particle wavefunction under center of mass
boosts. The time dependence induced by the unitary driving dynamics on
the particle-particle entanglement is algebraically akin to that
governing the change in time of the width parameter of the center of
mass wavepacket, but involves its own, different time scale. The two
time scales become even more distinct as the initial width
parameter of the center of mass wavepacket becomes appreciably smaller
than the bound state size parameter. In fact, we have seen that the
quadratic decrease, as $B\rightarrow 0$, of the characteristic time
$\tau_B$ for wavepacket spreading is not followed by the
characteristic time for the evolution of entanglement, which
approaches a finite limit involving properties of the two particle
bound state.

One must note also that the monotonic increase in time of both the
spreading of the initial width $B$ of the center of mass wavepacket and
of the linear entropy $\Delta$ used to measure the degree of
particle-particle entanglement hinge on the particular choice made for
the initial state. A still particular, but different choice leading to
also different behavior may be devised by taking advantage of the
time-reversible character of the overall unitary dynamics. In fact,
one may use as an alternate initial state $\tilde{\Psi}(k_1,k_2;0)$,
the time reversed counterpart of the state resulting from our adopted
initial state (\ref{instate}) after it evolves for a time $T$, i.e.

\[
\tilde{\Psi}(k_1,k_2;0)\equiv\Psi^*(-k_1,-k_2;T).
\]

\noindent Due to time-reversal invariance of the overall dynamics, the
time evolution of this state is given as $\tilde{\Psi}(k_1,k_2;t)=
\Psi^*(-k_1,-k_2;T-t)$, so that the time evolution from
$\Psi(k_1,k_2;0)$ to $\Psi(k_1,k_,T)$ is traced backwards in time. As
a result, the center of mass wavepacket will shrink and the linear
entropy will decrease as $t$ increases from $0$ to $T$. In particular,
if $\Psi(k_1,k_2;0)$ is a particle-factorizable initial state, the
linear entropy decreases smoothly from a non zero value to zero in the
finite time $T$, as in cases of the so called entanglement sudden
death~\cite{yueb}, but immediately and also smoothly rebounds to the
monotonic increase for subsequent times.

Finally, it is worth stressing that our simple model system provides
for instances of quantum states consisting of parts which are strongly
correlated dynamically but which are nevertheless {\it not entangled}.
This is in a way complementary to the situation treated by
Schr\"{o}dinger~\cite{schrod} in the wake of the Einstein, Podolski and
Rosen~\cite{einstein} argument, namely one in which entanglement
persists after interaction between the two parts has ceased. Both
cases point to a picture in which entanglement is ultimately a
quantum kinematical feature, even if circumstantially affected by
quantum dynamical processes.

\section*{Acknowledgments}

FRP thanks Marcio Corn\'{e}lio and Jonas Larson for useful discussions.
This work has been supported in part by CNPq, FAPESP, and the Swedish
Royal Council (Vetenskapr\aa det).

\section*{Appendix}

We give here, for completeness, a brief, elementary calculation
showing that the two reduced densities $\rho^{(i)}(k_1,k_2)$, $i=1,2$
have the same eigenvalues and that the wavefunction $\Psi(k_1,k_2)$
can in fact be written in the form shown in eq. (\ref{schmidt}), cf.
ref. \cite{vonNeumann}, Chap. VI, section 2. We omit the time
variable, which is not relevant here.

Note first that the set of eigenfunctions of $\rho^{(1)}$ can be
extended to a complete orthonormal set $\{\chi^{(1)}_n(k_1)\}$ in the
Hilbert space of particle 1 by including eigenfunctions with
eigenvalue zero.  Consider, in addition, a complete orthonormal set
$\{u_\nu(k_2)\}$ in the Hilbert space of particle 2. Then, for each of
the $\chi^{(1)}_n$ define

\[
\tilde{\chi}^{(2)}_n(k_2)=\sum_\nu u_\nu(k_2)(\chi^{(1)}_n u_\nu,\Psi),
\]

\noindent the bracket being the scalar product in product space. This
may be checked to be an eigenfunction of $\rho^{(2)}$. In fact, write
this reduced density as

\[
\rho^{(2)}(k_2,k_2')=\sum_n\int dk_1\int dk_1'\chi^{(1)*}_n(k_1)
\Psi(k_1,k_2)\Psi^*(k_1',k_2')\chi^{(1)}_n(k_1')
\]

\noindent so that 

\begin{eqnarray*}
\int dk_2'\rho^{(2)}(k_2,k_2')\tilde{\chi}^{(2)}_n(k_2')&=&\sum_{n'}
\int dk_1\chi^{(1)*}_{n'}(k_1)\Psi(k_1,k_2)\sum_\nu(\Psi,
\chi^{(1)}_{n'}u_\nu)(\chi^{(1)}_n u_\nu,\Psi)= \\ 
&=&\lambda^{(1)}_n\tilde{\chi}^{(2)}_n(k_2). 
\end{eqnarray*}

\noindent The sum over $\nu$ in the last expression involves the
reduced density $\rho^{(1)}$ and is in fact just $\lambda^{(1)}_n
\delta_{nn'}$, and the integral over $k_1$ yields the function
$\tilde{\chi}^{(2)}_n(k_2)$. This shows then that this function is an
eigenfunction of the reduced density $\rho^{(2)}$ with the same
eigenvalue as its parent eigenfunction $\chi^{(1)}_n(k_1)$ of
$\rho^{(1)}$. From now on we may thus omit the eigenvalue
superscripts.

Next check the norms of the functions $\tilde{\chi}^{(2)}_n$. This is
easily done by evaluating

\[
(\tilde{\chi}^{(2)}_{n'},\tilde{\chi}^{(2)}_n)=\sum_\nu
(\Psi,{\chi}^{(1)}_{n'}u_\nu)({\chi}^{(1)}_{n}u_\nu,\Psi)=
\lambda_n\delta_{nn'},
\]

\noindent showing that the functions $\chi^{(2)}_n(k_2)=
\tilde{\chi}^{(2)}_n(k_2)/\sqrt{\lambda_n}$ constitute an orthonormal
set. On the other hand, $\Psi(k_1,k_2)$ may be expanded in the product
base $\{\chi^{(1)}_n(k_1) u_\nu(k_2)\}$:

\begin{eqnarray*}
\Psi(k_1,k_2)&=&\sum_n\sum_\nu\chi^{(1)}_n(k_1) u_\nu(k_2)(\chi^{(1)}_n
u_\nu,\Psi)= \\ &=& \sum_n\chi^{(1)}_n(k_1)\tilde{\chi}^{(2)}_n(k_2)=
\sum_n\sqrt{\lambda_n}\chi^{(1)}_n(k_1)\chi^{(2)}_n(k_2)
\end{eqnarray*}

\noindent which establishes eq. (\ref{schmidt}).

%\section*{References}
%\begin{enumerate}


\begin{thebibliography}{99}

\bibitem{schrod} E. Schr\"{o}dinger, ``Discussion of Probability Relations between Separated Systems'', {\it Proceedings of the Cambridge
    Philosophical Society} {\bf 31}, 555-563 (1935).

\bibitem{einstein} A. Einstein, B. Podolsky, and N. Rosen, ``Can Quantum-Mechanical Description of Physical Reality Be Considered Complete?'', {\it Phys. Rev.} {\bf 47}, 777-780 (1935).

\bibitem{bell} J. Bell, {\it Speakable and Unspeakable in Quantum
mechanics}, Cambridge University Press, 1987.

\bibitem{aspect} A. Aspect, P. Grangier, and G. Roger, ``Experimental Tests of Realistic Local Theories via Bell's Theorem'', {\it
    Phys. Rev. Lett.} {\bf 47}, 460-463 (1981).

\bibitem{mermin} N. D. Mermin, ``Bringing home the atomic world: Quantum mysteries for anybody'', {\it Am. J. Phys.} {\bf 49}, 940-943 (1981).

\bibitem{odradeks} R. Horodecki, P. Horodecki, M. Horodecki, and
  K. Horodecki, ``Quantum entanglement'', {\it Rev. Mod. Phys.} {\bf 81}, 865-942 (2009).

\bibitem{niels} M. A. Nielsen and I. L. Chuang {\it Quantum
    Computation and Quantum Information}, Cambridge University Press
  2000. 

\bibitem{presk} John Preskill {\it Lecture Notes for Physics 229:
    Quantum Information and Computation}, September, 1998.

\bibitem{mint} F. Mintert, A. R. R. Carvalho, M. Kus, and
  A. Buchleitner, ``Measures and dynamics of entangled states'', {\it Phys. Reports} {\bf 415}, 207-259 (2005).

\bibitem{zurek} W. H. Zurek, ``Decoherence and the Transition from
    Quantum to Classical - Revisited'', {\it arXiv:quant-ph/0306072} (2003);
  {\it Rev. Mod. Phys.} {\bf 75}, 715-775 (2003).

\bibitem{omnes} R. Omnès, {\it The Interpretation of Quantum
    Mechanics}, Princeton University Press, 1994.

\bibitem{yueb} Ting Yu and J. H. Eberly, ``Finite-Time Disentanglement via Spontaneous Emission'', {\it Phys. Rev. Lett.} {\bf
    93}, 140404 (2004); ``Sudden Death of Entanglement'', {\it Science} {\bf 323}, 598-601 (2009).

\bibitem{rio} M. P. Almeida, F. de Melo, M. Hor-Meyll, A. Salles,
  S. P. Walborn, P. H. Souto Ribeiro, and L. Davidovich, ``Environment-Induced Sudden Death of Entanglement'', {\it Science}
  {\bf 316}, 579-582 (2007).

\bibitem{lljn} Z.-J. Li, J.-Q. Li, Y.-H. Jin, and Y.-H. Nie, ``Time evolution and transfer of entanglement between an isolated atom 
and a Jaynes-Cummings atom'', {\it J. Phys. B: At. Mol. Opt. Phys.} {\bf 40}, 3401-3411 (2007). 

\bibitem{vonNeumann} J. von Neumann, {\it Mathematical Foundations of
    Quantum Mechanics}, Princeton University Press, 1955, Chap. V,
  section 3.

\bibitem{Falk} H. Falk, ``Inequalities of J. W. Gibbs'', {\it Am. J. Phys.} {\bf 38}, 858-869 (1970).

\bibitem{PRIT} P. Tommasini, E. Timmermans, and A. F. R. de Toledo
  Piza, ``The hydrogen atom as an entangled electron-proton system'', {\it Am. J. Phys.} {\bf 66}, 881-886 (1998).

\end{thebibliography}
\end{document}